\documentclass{astro}

\usepackage{epsf}

\newcommand{\md}{\mbox{d}}

\begin{document}

   \thesaurus{03               
              (11.01.2;        
               11.02.2 Mkn 501 
               11.10.1)}       
              
   \titlerunning{Implications of a possible 23 day periodicity in Mkn~501}
       
   \title{Implications of a possible 23 day periodicity for binary black hole
          models in Mkn~501}

   \author{F. M. Rieger \& K. Mannheim}

   \offprints{F.M. Rieger (frieger@uni-sw.gwdg.de)}

   \institute{Universit\"ats-Sternwarte, Geismarlandstr. 11, 
              D-37083 G\"ottingen, Germany}

   \date{Received 4 May 2000/\, Accepted 11 May 2000}

   \maketitle

   \begin{abstract}       
      We investigate the implications of a massive binary system in the 
      centre of the gamma-ray blazar Mkn~501 and show that the periodical 
      behaviour recently observed in the TeV and X-ray lightcurves may 
      possibly be related to the orbital motion of the relativistic 
      jet emerging from the less massive black hole. For the special 
      relativistic jet properties inferred from emission models, we 
      derive an intrinsic orbital period of $(6-14)$ yrs and a 
      centre-of-mass distance of $(2.0-3.5) \times 10^{16}$ cm. 
      If the binary is very close with a separation of the order of 
      that for which gravitational radiation becomes dominant, we find
      a maximum primary mass of $\sim 10^8 M_{\sun}$ and a corresponding
      secondary mass in the range of $\sim (4-42)\times 10^6 M_{\sun}$ 
      depending on the intrinsic jet properties. 
      Such values are in line with the black hole masses expected from 
      merger scenarios.

      \keywords{Galaxies: active -- BL Lacertae objects:
       individual: Mkn~501 -- Galaxies: jets}
   \end{abstract}

\section{Introduction}
    Binary black hole systems (BBHSs) are expected to be common in the 
    universe as a result of mergers between galaxies. In the underlying 
    picture for the morphological evolution, galaxies were formed as part 
    of a hierarchical clustering process (e.g. White~1997). Giant elliptical 
    galaxies, such as the host galaxy of Mkn~501, appear to be the products 
    of mergers between spiral galaxies (cf. Fritze v.- Alvensleben~1996).
    Since the brightest galaxies generally seem to contain massive black 
    holes in their nuclei (e.g. Rees~1984; Kormendy \& Richstone~1995; 
    Ho~1998; Magorrian et al.~1998; Richstone et al.~1998), merging would 
    naturally lead to the formation of massive BBHS (Begelman et al.~1980, 
    abbreviated: BBR~80; Rees~1994; Artymowicz~1998; Richstone~1998).
    The so formed binary black hole is expected to spent most of its time 
    at a separation of $\sim 0.1 - 1$ pc for masses of $10^8\,M_{\sun}$
    (BBR~80). However if the binary looses further angular momentum e.g. by 
    slingshot interaction with new stars from subsequent merging events 
    (Roos~1988; Roos et al.~1993), infall of gas (BBR~80; cf. also 
    Gould \& Rix~2000) or by interactions with an accretion disk 
    (Ivanov et al.~1999), gravitational radiation will eventually become 
    important and the binary evolution could proceed rapidly to 
    coalescence.

    Up to now, several phenomena have been attributed to BBHSs: 
    e.g. misalignment (cf. Conway \& Wrobel~1995), precession 
    (BBR~80) or wiggling of jets, where the latter is supposed to be 
    induced by the orbital motion (Kaastra \& Roos~1992; Roos et al.~1993). 
    Periodic outburst activity in the quasar OJ~287, has commonly been 
    related to a BBHS and is thought to arise due to tidal 
    perturbation (Sillanp\"a\"a et al.~1988) or due to one black 
    hole crossing the accretion disk of the other (Letho \& Valtonen~1996). 
    Other BBHS scenarios assume a pair of bent jets (Villata et al.~1998) or 
    the precession of the disk under the gravitational torque (Katz~1997). 
    
    In the particular case of Mkn~501, the complex morphology of its radio 
    jet and the peculiar behaviour of its spectral energy distribution 
    (SED) have prompted elaborate models relating these properties to a BBHS: 
    Conway \& Wrobel~(1995), for example, have proposed a saturated helix model 
    in order to explain the misalignement of the radio jet on parsec and kiloparsec 
    scales. 
    Villata \& Raiteri~(1999) have argued that the X-ray variations in the 
    SED of Mkn~501 might be due solely to the changing orientation of a 
    helical synchrotron emitting jet in a close BBHS.
    The recent discovery of periodicity in the TeV and X-ray fluxes believed
    to be associated with moving features in the jet of Mkn~501 might add 
    another aspect for assessing the relevance of a BBHS in this galaxy.
    
    Mkn~501 is one of at least four active galactic nuclei which have been 
    detected at TeV energies (for review, see Catanese \& Weekes~1999). 
    Being the second closed among these with a redshift of $z=0.034$, 
    Mkn~501 has been historically classified as an X-ray selected BL Lac 
    object showing virtually no emission lines, and is hosted by the 
    elliptical galaxy UGC 10599 (Stickel at al.~1993). 
    As a BL Lac object, Mkn~501 belongs to the blazar class of AGN which
    are thought to have relativistic jets oriented at a small viewing angle, 
    thus yielding a strong Doppler enhancement of the observed flux.
 
    At the beginning of 1997, Mkn~501 had suddenly undergone a phase of high 
    activity becoming the brightest source in the sky at TeV energies. 
    Subsequent multiwavelength campaigns revealed a variable, two component 
    SED with a low energy part extending up to $100$ keV (Pian et al.~1998) 
    and a high energy part which extends at least up to 20 TeV 
    (Samuelson et al.~1998; Konopelko~1999).  
    During this activity phase, particular types of variability have been 
    observed (e.g. Protheroe et al. 1998), consisting of flaring episodes 
    of several days and additional intraday-variabilities. 
    While the TeV and X-rays variations seem to be well correlated, the
    evidence for correlations with the optical U-band appears to be rather 
    weak (e.g. Catanese et al~1997; Djannati-Atai et al.~1999; 
    Aharonian et al.~1999). 
    One of the most fascinating features is the observed periodicity in the
    TeV region with a period in the range of $(23-26)$ days, which has been
    found in the data taken by several Cherenkov telescopes (see Protheroe 
    et al.~1998; Hayashida et al.~1998). 
    Additionally, a recently performed analysis of RXTE-ASM X-ray data from 
    April to July 1997 also seems to support a periodicity of $\sim 23$ days 
    (Kranich et al.~1999; Nishikawa et al.~1999). 

    Here, we speculate on the possibility that this periodicity in the flaring
    state arises due to the orbital motion of the relativistic jet in a BBHS  
    where the nonthermal radiation is emitted by a relativistic jet which emerges 
    from the less massive black hole, the periodicity thus being due mainly 
    to geometrical origin (i.e. Doppler-shifted modulation). 

\section{On the influence of a BBHS in Mkn~501}    
    \subsection{Doppler-shifted flux modulation}
    Consider a simple binary model for Mkn~501, where the binary orbit is
    is assumed to be circular because dynamical friction between two parent 
    galaxies during the merger might ensure that the initial eccentricity of 
    the resulting binary is small (e.g. Polnarev \& Rees~1994). 
    Using Kepler's third law, the intrinsic orbital frequency $\Omega_k$ of 
    a binary with separation $d$ is given by
    \begin{equation}\label{kepler}
    \Omega_k=\frac{\sqrt{G\,(m+M)}}{d^{3/2}}\,,
    \end{equation}where $m$ and $M$ denotes the mass of the smaller 
    and the larger hole, respectively, and $G$ is the gravitational 
    constant.

    Let us further assume that the observed jet is formed by the less
    massive black hole and that the nonthermal X-ray and $\gamma$-ray 
    radiation in the flaring state is emitted by a relativistic emission 
    region (e.g. knot, blob, shock) which propagates outwards from the core 
    along the jet with gamma factor $\gamma_{b}$.  
    Owing to the (non-relativistic) orbital motion, the true trajectory of 
    the knot is a long stretched helix.
    The modulation of the emission then occurs as a consequence of the 
    slight variation of the inclination angle $i$ due to the orbital 
    $\phi$-component of the velocity field of the knot.
    This observed flux modulation by Doppler boosting is well-known and
    can be written in the form
    \begin{equation}\label{modulo}
    S(\nu)=\delta^3\,S'(\nu')=\delta^{3+\alpha}\,S'(\nu),
    \end{equation}
    where $S'$ is the spectral flux density measured in the comoving frame,
    $\delta(t)=1/(\gamma_{b}[1-\beta_b\,\cos\theta(t)])$ the Doppler 
    factor, $\theta(t)$ is the actual angle between the velocity 
    $\vec \beta_b =\dot{\vec x}_b(t)/c$ of the emission region and the direction 
    of the observer, and where the final equality holds if the source has a 
    spectral index $\alpha$.

    Due to the orbital motion around the center-of-mass, the Doppler 
    factor for the emission region is a periodical function of time. 
    In the simplest case where the angle between the jet axis and the
    direction of the total angular momentum of the binary is assumed to
    be zero (e.g. neglecting any kind of precessional motion) the Doppler
    factor may be written as
    \begin{equation}\label{doppler}    
    \delta=\frac{\sqrt{1-(v_z^2 + \Omega_k^2 R^2)/c^2}}{1-
    (v_z \cos i-\Omega_k R\,\sin i \sin \Omega_k\,t)/c}\,,
    \end{equation}with $R=M\,d/(m+M)$ being the centre-of-mass distance,
    $v_z$ the outflow velocity in the direction of the total angular 
    momentum, $i$ the inclination between the jet axis and the line of 
    sight and $c$ the velocity of light. Obviously, the Doppler factor 
    becomes maximal for $t=0.75\,P_k$ and minimal for $t=0.25\,P_k$, 
    where $P_k=2\,\pi/\Omega_k$ denotes the keplerian period.

    From the TeV flux ratio of $f \sim 8$ between the maximum 
    and the minimum state during the observation (cf. Protheroe et 
    al.~1998; Hayashida et al.~1998, Aharonian et al.~1999) and the 
    assumption that the periodicity arise in the main due to geometrical 
    origin, we now obtain the condition $\delta_{\rm max}/\delta_{\rm min} 
    \simeq f^{1/(3+\alpha)}$ (see Eq.~(\ref{modulo})). Consequently, by using 
    Eq.~(\ref{doppler}) one finds
    \begin{equation}\label{ratio}
    \Omega_k\,R=\frac{f^{1/(3+\alpha)}-1}{f^{1/(3+\alpha)}+1}\,
    \left(\frac{1}{\sin i}-\frac{v_z}{c}\,\cot i\right)\,c\,.
    \end{equation}
    
    For a source region which moves in the time interval $dt$ from point 
    $A$ to point $B$ with relativistic velocity $v_z$ and at an angle 
    $\psi$ to the line of sight, the observed difference in arrival 
    times for radiation emitted at $A$ and $B$ is generally given by
    $dt_{\rm obs} = dt - dt \,(v_z /c)\,\cos\psi$, thus leading to 
    a shortening of the observed time interval. 
    Along this line of argument, one may easily derive that the observer 
    in the model presented here will only perceive a strongly shortened 
    period, i.e. the observed period $P_{\rm obs}$ is related to the 
    intrinsic period $P_k$ by (cf. also Camenzind \& Krockenberger~1992, 
    Roland et al.~1994)
    \begin{equation}
    P_{\rm obs}=(1+z)\,\int_0^{P_k}(1-\beta_b\,\cos\theta(t))\,\md t\,.
    \end{equation} 
    Performing the integration, one immediately arrives at 
    \begin{equation}\label{time}
    P_{\rm obs}=(1+z)\,(1-\frac{v_z}{c}\cos i)\, P_k\,.
    \end{equation}
    From the theoretical point of view, relativistic blazar jets are thought 
    to be oriented at a small viewing angle. Current emission models favour 
    an inclination angle $i \simeq 1/\gamma_b$ (Spada~1999; cf. also Chiaberge et 
    al.~2000) with typical bulk Lorentz factors in the range $10-15$ (e.g. Mannheim 
    et al.~1996; Hillas~1999; Spada et al.~1999). For such values and by using an 
    observed period of $23$ days and a characteristic outflow velocity of 
    $v_z/c \simeq (1-1/\gamma_b^2)^{0.5}$, Eq.~(\ref{time}) results in an intrinsic 
    period of $P_k = (6-14)$ yrs.

    Combining Eq.~(\ref{time}) and Eq.~(\ref{ratio}) we may also derive an
    expression for the the centre-of-mass distance
    \begin{equation}\label{distance}
    R=\frac{P_{\rm obs}}{2\,\pi\,(1+z)}\,
      \frac{f^{1/(3+\alpha)}-1}{f^{1/(3+\alpha)}+1}\,\frac{c}{\sin i}\,.
    \end{equation} Given the observed period and the spectral index, 
    Eq.~(\ref{distance}) only depends on the inclination angle. Accordingly, 
    for an observed period of $23$ days, a ratio $f=8$ and a TeV spectral index 
    of $\alpha \simeq (1.2-1.7)$ (cf. Aharonian et al.~1999), one gets 
    $R \simeq (2.0-3.5)\times 10^{16}$ cm using the inclination values above.

    By inserting Eq.~(\ref{kepler}) in Eq.~(\ref{distance}), the appropriate 
    binary mass ratio is given by
    \begin{eqnarray}\label{massratio}
    \frac{M}{(m+M)^{2/3}} & = & \frac{P_{\rm obs}^{1/3}}{(2\,\pi\,[1+z]\,
                    G)^{1/3}}\,\frac{c}{\sin i}\,
                    \nonumber \\            
                    & &\times \frac{f^{1/(3+\alpha)}-1}{f^{1/(3+\alpha)}+1}\,
                        (1-\frac{v_z}{c}\,\cos i)^{2/3}\,.
    \end{eqnarray}
    For a secondary mass in the range of $(10^6-10^8)\, M_{\sun}$ the 
    required primary masses are calculated in Fig.~\ref{mass} 
    (see the curves $K(10),K(15)$) for two different inclination angles
    and $f=8$ yielding primary masses of the order of $10^8\,M_{\sun}$.
    
    \subsection{A gravitational constraint on the binary separation}
\begin{figure}[htb]
       \vspace{0cm}
       \begin{center}
         \epsfxsize=8.5cm         
       \mbox{\epsffile{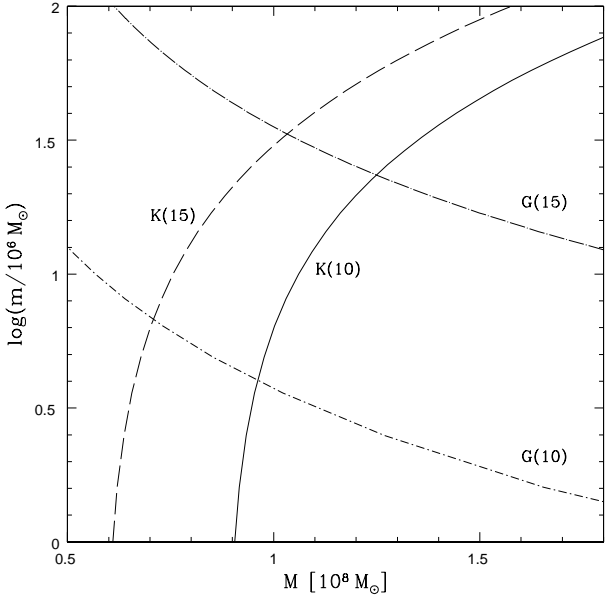}}
       \end{center}
       \vspace{-0.5cm} 
       \caption{Required mass dependence for a BBHS in Mkn 501. 
        The solid [$K(10)$] and long dashed [$K(15)$] curve are
        given by the Doppler condition Eq.~(\ref{massratio}) for 
        inclination angles $i=1/\gamma_b$ with $\gamma_b=10$ and $15$, 
        respectively. The curves $G(1/i)$ are given by the condition that 
        the current binary separation equals the gravitational distance 
        $d_g$, see Eq.~(\ref{grav}), thus yielding upper limits on the 
        allowed binary masses. A TeV spectral index of $1.2$ has been applied
        for the calculation.}
        \vspace{0cm}
       \label{mass}
   \end{figure}

    Observationally, BL Lacs are in general less luminous radio sources, 
    showing a lack of strong optical emission lines and little signs of 
    cosmological evolution (cf. Bade et al.~1998; Cavaliere \& Malquori~1999). 
    Celotti et al.~(1998) have suggested that BL Lac objects correspond
    to the final evolutionary stage of sources accreting at low radiative
    efficiencies (i.e. a dormant black hole system), which seems to be 
    supported by HST observation indicating that the less luminous AGN 
    stages occur after the original quasar has dimmed (Bahcall et al.~1994).
    Recently, Villata \& Raiteri~(1999) have argued that BL Lacs represent
    advanced and close BBHS with a decreased mass accretion rate, the binary 
    separation in the case of Mkn~501 being of the order of that for which 
    gravitational radiation becomes dominant. 
    Thus, we might set an upper limit on the allowed binary masses in Mkn~501 
    by assuming that the current separation equals the gravitational 
    separation $d_g$, i.e. the position where the gas dynamical time scale 
    is balanced by the time scale for gravitational energy losses (BBR~80). 
    Gas, which may be constantly supplied for example by tidal interaction 
    between galaxies (cf. Heidt~1999) and accreted onto the more massive 
    black hole, will cause the binary separation to shrink on a time scale 
    $t_{\rm gas} \simeq M\, (1 M_{\sun}\,\rm{yr^{-1}}/\,\dot{M})$, 
    with $\dot{M}$ the accretion rate (BBR~80). For a simple estimate 
    let us assume that during the optical bright QSO phase mass accretion 
    occurs at about the Eddington limit. The phase of nuclear activity seems 
    to be rather short with a typical duration of a few times $10^7$ yrs 
    (Haehnelt et al.~1998; Richstone et al.~1998). 
    In particular, for a duration of nuclear activity of the order of the 
    salpeter lifetime $t_{\rm s}=\epsilon \sigma_T c/4\pi G m_p = 
    4.5\,\epsilon_{0.1}\times 10^7$ yrs and for a primary black hole mass 
    of $\sim 10^8\,M_{\sun}$, gas infall rates of $\sim 2 \,M_{\sun} 
    \rm{yr}^{-1}$ are required to sustain the Eddington luminosity, 
    using a canonical $10\%$ efficiency.
    On the other hand, assuming a circular orbit, the time scale for 
    gravitational radiation is given by $\tau_{\rm grav} = 
    6.3 \times 10^{4}\,d_{16}^{\,4}/(M_8\,m_8\,[m_8+M_8])$ yrs, where  
    the distance and the masses are expressed in units of $10^{16}\,{\rm cm}$ 
    and $10^8\,M_{\sun}$, respectively.  
    Thus, by equating the gas dynamical time scale $t_{\rm gas}$ with 
    $\tau_{\rm grav}$, the separation at which gravitational radiation 
    becomes dominant may be written as 
    \begin{eqnarray}\label{gravrad}
    d_{g_{16}}=6.3 \, M_8^{1/2}\,m_8^{1/4}\,(m_8+M_8)^{1/4}
        \left(\frac{M_{\sun}\,{\rm yr}^{-1}}{\dot{M}}\right)^{1/4}\,.    
    \end{eqnarray}
    From Eq.~(\ref{gravrad}) and Eq.~(\ref{kepler}) we immediately
    get the relation
    \begin{eqnarray}\label{grav}
    \frac{M^{1/2} m^{1/4}}{(m+M)^{1/12}} & = &  1.29 \times 10^{22}\,
             \left(\frac{\dot{M}}{M_{\sun}\,{\rm yr}^{-1}}\right)^{1/4}
             \nonumber \\
             & & \times \frac{P_{\rm obs}^{2/3}}{(2\,\pi\,[1+z])^{2/3}\,
                 (1-\frac{v_z}{c}\,\cos i )^{2/3}}\,.
    \end{eqnarray}
    This mass dependence is illustrated in Fig.~\ref{mass} (curves $G$). 
    The respective upper limit is given by the point of intersection
    with the relevant curve $K$. For example, applying $\alpha =1.2$ and 
    using $i=1/10$, we have a maximum secondary mass $m \simeq 4 \times 
    10^6\,M_{\sun}$ and a corresponding primary mass of $M \simeq  
    10^8\,M_{\sun}$(cf. also Table~\ref{tab}).
    The masses shown in Fig.~\ref{mass} are in a reasonable range for 
    ellipticals. Masses of the order of one million solar masses for the
    companion black hole appear to be in agreement with the concept that 
    the galaxy swallowed in the merger process was a minor spiral galaxy. 
    On the other hand, the host galaxy of Mkn~501 seems to belong to these 
    classes of ellipticals which have black holes in the centers of at least 
    a few hundred million solar masses. Therefore, the binary scenario for 
    Mkn~501 seems not unlikely.

\section{Discussion}
    
     In this paper we have suggested that the periodicity in the flaring state 
     observed in Mkn~501 might be caused by the orbital motion of the jet in a close 
     BBHS. Applying a simple toy-model we have shown that the BBHS may have a period  
     of $\sim (6-14)$ yrs and a centre-of-mass distance of $\sim (2.0-3.5) 
     \times10^{16}$ cm. If one assumes that this separation corresponds to  
     the distance at which gravitational radiation becomes important, several 
     upper limits for the binary masses may be derived. These mass ranges, which
     are shown in Table~\ref{tab} using an observed period $P_{\rm obs}=23\,
     {\rm days}$ seem to be in line with the expectations from merger scenarios
     and the suggestions made by Villata \& Raiteri~(1999).
      
  \begin{table}[hbt]
  \vspace{0cm}
  \begin{center}
  \begin{tabular}{|l@{\hspace{0.7cm}}|l@{\hspace{0.7cm}}
                |l@{\hspace{0.7cm}}|}\hline
  $i=1/\gamma_b$  &  $1/10$       &  $1/15$  \\ \hline \hline
  $m\,[10^8\,M_{\sun}]$   & 0.040 \quad (0.061) & 0.33 \quad (0.42) \\ \hline
  $M\,[10^8\,M_{\sun}]$   & 0.96 \quad (0.75)   & 1.03 \quad (0.91) \\ \hline
  $d\,[10^{16}\,\rm{cm}]$ & 2.40 \quad (2.24)   & 4.57 \quad (4.54)  \  \\ \hline
  $P_k\,[\rm{yrs}]$        & 6.10                & 13.7    \\ \hline 
  $\tau_{\rm grav}\,[10^7\,\rm{yrs}]$ & 5.46 \quad (4.25) & 5.86 \quad (5.18) \\ \hline
   $P_p\,[10^4\,\rm{yrs}]$  & 0.52 \quad (0.39) & 3.03 \quad (2.93)  \\ \hline 
  \end{tabular}
  \end{center}
  \vspace{0cm}
  \caption{Maximum binary masses, separation $d$, intrinsic orbital period 
           $P_k$, gravitational lifetime $\tau_{\rm grav}$ and precessional 
           period $P_p$ for inclination angles $i$, accretion rate $\dot{M}
           =2 \,M_{\sun}/{\rm yr}$ and spectral index $\alpha=1.2\,\,(1.7)$.}
  \label{tab}
  \vspace{0cm}
  \end{table}
     The TeV observations indicate that we may have $N \leq 6$ for the number $N$ of 
     periodic oscillations (cf. Aharonian et al.~1999; Catanese \& Weekes~1999;
     Quinn et al.~1999), which results in a required propagation length for 
     the emitting component of $l_z=N\,P_k\,v_z \simeq 11-26$ pc. Thus, for the 
     projected length at the position of Mkn~501 one finds $l_p \simeq 1.4-2.1$ mas 
     for $i=(1/15)-(1/10)$ rad. Remarkably, the jet of Mkn~501 bends dramatically 
     at about $3$ mas from the core (Marscher~1999).
     Hence, a change in the jet parameters might be the reason for the 
     termination of the observed periodicity. 

     For the proposed model to be valid, the jet has to be perfectly 
     collimated with an intrinsic opening angle of less then $\arctan(d/l_z) 
     \sim 0.05^{\circ}$. Such values are indeed expected in scenarios for 
     the formation and collimation of magnetized BL Lac jets (cf. 
     Camenzind \& Krockenberger~1992; Appl \& Camenzind~1993;
     Schramm et al.~1993). At first sight, such a cylindrical jet structure 
     seems to be at least $\sim 20$ times more collimated than the jet seen on  
     VLBA maps (cf. Marscher~1999). However, there is evidence for an at 
     least two-component jet structure in Mkn~501 suggesting an inner spine with 
     transverse magnetic field and an envelope with longitudinal magnetic field 
     (Aaron~1999; Marscher~1999), the polarization properties of the inner spine 
     strongly supporting shocked-jet models (cf. Attridge et al.~1999).
     In fact, our model requires that the high energy emission originates 
     in a channel along the jet axis as in two-fluid models (e.g. Sol et al.~1989, 
     Roland et al.~1994), the inner emission probably being self-absorbed on the VLBA 
     scale. Recent observations of radio jets indeed indicate a confinement of the 
     higher energy emission to a well-defined channel within a much more extended 
     radio emission (Bahcall et al.~1995; Perlman et al.~1999; Swain et al.~1999).
     The unification of BL Lacs and FR I objects may add another piece of evidence
     to such a jet configuration: in order to account for the observed spectral 
     properties an at least two-fold jet velocity structure seems to be required in 
     which a fast spine is surrounded by a slow (but still relativistic) layer 
     (Chiaberge et al.~2000). Support for such a possibility is positively provided 
     by numerical simulations (cf. Aloy et al.~2000; Frank et al.~2000).

     For a TeV flux ratio between minimum and maximum of $\sim 8$, the corresponding
     shift in the break frequency would be given by a factor of $\sim (1.5-1.6)$ while
     the X-ray flux ratio becomes $\sim (5-7) $ applying an hard X-ray spectral index 
     $\alpha \simeq 0.6-0.9$ (cf. Lamer \& Wagner~1998; Pian et al.~1998). 
     Such values seem to be consistent with BeppoSAX observations (Pian et al.~1998) 
     and may also be recovered, using a broken power law fit, in RXTE observations of 
     Mkn~501 (cf. Krawczynski et al.~2000, their Figs.1 and 2a).     
     Gamma-ray observations carried out by the CAT Telescope also reveal a
     shift of the maximum peak energy apparently in accordance with the 
     expectation above (Djannati-Atai et al.~1999).
     Small changes in the maximum electron Lorentz factor or the magnetic field
     along the trajectory of the emission region may further add to flux 
     variations. If there is indeed an additional flux contribution, e.g. low energy 
     emission from the layer, a stationary component comparable to the observed 
     infared-optical flux (e.g. Pian et al.~1998; cf. also Kataoka et al.~1999) 
     or an additional component responsible for the soft X-ray emission 
     (e.g. Lamer \& Wagner~1998; Wagner et al.~1999), the amplitude of the 
     Doppler modulation may decrease to lower frequencies.  
      
     In the simple model presented above, we have not yet considered the 
     influence of jet precession due to gravitomagnetic and geodetic origin 
     with a period $P_p \simeq 580 \times \,d_{\,16}^{\,5/2}\,(M_8+m_8)^{1/2}\,
     m_8^{-1}\,M_8^{-1} (1+3 M/4 m)^{-1} \rm{yrs}$ (cf. Thorne et al.~1986). 
     Since this driving period is much larger than the orbital period 
     (cf. Table~\ref{tab}), a precessional modulation should be negligible during 
     a few revolutions. Interestingly, a precessional period of $\sim 10^4$ yrs 
     agrees with the driving frequency found by Conway \& Wrobel~(1995) in order
     to explain the misalignement of the radio jet in Mkn~501 on parsec and 
     kiloparsec scale (see also Villata \& Raiteri~1999).
     If the binary hypothesis is correct, the observable period should 
     remain similar during different outburst phases unless there is a change 
     in the general jet properties. For example, an increase in the observed 
     period should then be accompanied by a decrease in the bulk Lorentz factor
     or, on larger time scales, by an increase of the inclination angle due to the 
     jet precession (cf. also Eq.~\ref{time}).

\begin{acknowledgements}
      We would like to thank an anonymous referee for useful suggestions.
      K.M. acknowledges support from a Heisenberg-Fellowship and F.M.R. 
      from DFG Ma 1545/2-1.
\end{acknowledgements}

\end{document}